\documentclass[aps,twocolumn,pra,showpacs,10pt]{revtex4-1}
\usepackage[T1]{fontenc}
\usepackage{graphicx,amsmath,amssymb,xspace}
\usepackage{color}

\newcommand{\ket}[1]{| #1 \rangle}
\newcommand{\bra}[1]{\langle #1 |}
\newcommand{\Tr}{\mathrm{Tr}}

\newcommand{\C}{\mathbb{C}}

\newcommand{\kb}[1]{\ket{#1} \bra{#1}}

\newcommand{\thmref}[1]{\hyperref[#1]{{Theorem~\ref*{#1}}}}
\newcommand{\lemref}[1]{\hyperref[#1]{{Lemma~\ref*{#1}}}}
\newcommand{\corref}[1]{\hyperref[#1]{{Corollary~\ref*{#1}}}}
\newcommand{\eqnref}[1]{\hyperref[#1]{{Equation~(\ref*{#1})}}}
\newcommand{\claimref}[1]{\hyperref[#1]{{Claim~\ref*{#1}}}}
\newcommand{\remarkref}[1]{\hyperref[#1]{{Remark~\ref*{#1}}}}
\newcommand{\propref}[1]{\hyperref[#1]{{Proposition~\ref*{#1}}}}
\newcommand{\factref}[1]{\hyperref[#1]{{Fact~\ref*{#1}}}}
\newcommand{\defref}[1]{\hyperref[#1]{{Definition~\ref*{#1}}}}
\newcommand{\exampleref}[1]{\hyperref[#1]{{Example~\ref*{#1}}}}
\newcommand{\hypref}[1]{\hyperref[#1]{{Hypothesis~\ref*{#1}}}}
\newcommand{\secref}[1]{\hyperref[#1]{{Section~\ref*{#1}}}}
\newcommand{\chapref}[1]{\hyperref[#1]{{Chapter~\ref*{#1}}}}
\newcommand{\apref}[1]{\hyperref[#1]{{Appendix~\ref*{#1}}}}
\newcommand\rank{\mbox{\tt {rank}}\xspace}

\def\be{\begin{equation}}
\def\ee{\end{equation}}

\newcommand{\comment}[1]{{}}

\begin{document}

\title{\vspace{-1cm}Device-independent dimension tests in the prepare-and-measure scenario}

\author{Jamie Sikora$^{1,2}$, Antonios Varvitsiotis$^{1,2,3}$, and Zhaohui Wei$^{1,2,3}$}
\email{Email: weizhaohui@gmail.com} \affiliation{$^{1}$Centre for
Quantum Technologies, National University of Singapore,
Singapore\\$^{2}$MajuLab, CNRS-UNS-NUS-NTU International Joint
Research Unit, UMI 3654, Singapore\\$^{3}$School of Physical and
Mathematical Sciences, Nanyang Technological University, Singapore}

\begin{abstract}
Analyzing the dimension of an unknown quantum system in a
device-independent manner, i.e., using only the measurement
statistics, is a fundamental task in quantum physics and quantum
information theory. In this paper, we consider this problem in the
prepare-and-measure scenario. {Specifically, we provide a
 lower bound on the dimension of
the prepared quantum systems which is a function that only depends
on the measurement statistics.  Furthermore, we show that our bound
performs well on several examples. {In particular}, we show that
our bound provides new insights
{into}
 the notion of dimension witness,
and  {we also use it to show}
 that the sets of {restricted-dimensional} 
prepare-and-measure correlations are not always convex.}

\end{abstract}

\maketitle

{In the device-independent paradigm one tries to understand the
properties of an unknown (classical or quantum) system  based only
on the correlations resulting from measurements performed on the
system~\cite{BPA+08,WCD08,VSW15,BCP14,GBHA10,BQB14,HGM+12,ABCB12,LBL+15,BNV13,CBRS16,GBS16,CKKS16,CBB15}.
In this work we consider the problem of  lower bounding the
 dimension of a uncharacterized quantum system in a
device-independent  manner. This problem is quite interesting from
the viewpoints of {both physics and} quantum computation, and has
attracted much
attention~\cite{BPA+08,WCD08,VSW15,BCP14,GBHA10,BQB14,HGM+12,ABCB12,BNV13,CBRS16}.
Indeed, the dimension of a quantum system is {a fundamental physical
property, and is also} widely regarded as a valuable {computational}
resource, {as} one always tries to implement an algorithm or
protocol with the smallest dimension~{possible}}.

This question  was first considered in {the} Bell
{scenario}~\cite{BPA+08,WCD08,VSW15,BCP14} where a quantum system is
shared by two {parties},
 each performing local measurements on their
own subsystems, illustrated {in}
 FIG. 1(a). The {corresponding {set of} measurement statistics
is called a \emph{Bell correlation}}. Then the task is, for a given
Bell correlation, to lower bound the dimension of the underlying
quantum system. In \cite{BPA+08}, the {dimension witness approach}
was introduced to address this problem.
 {\em A (linear) $d$-dimensional
witness} is defined as a hyperplane that contains all Bell
correlations that can be generated using $d$-dimensional quantum
systems in one of its halfspaces. However, though providing very
strong and intuitive physical insights, dimension witnesses {can}
suffer from two apparent drawbacks. First, {dimension witnesses do
not {always} give a lower bound on the dimension of the underlying
quantum system as a direct function of the correlation data.}
Second, {identifying dimension witnesses amounts to characterizing}
the complicated structure of quantum correlations with restricted
dimensions, which is often a very challenging task.

{Another approach was recently introduced in an attempt to overcome
these difficulties.
  Specifically,  a new  lower bound on the dimension of a quantum system needed to generate
  a Bell correlation was given in \cite{VSW15}.  This bound is easy to calculate as it is a simple
  function of the Bell correlation and is tight in many  cases. }

The dimension witness approach was later generalized to the
prepare-and-measure (PM) scenario which is simpler and more
general~\cite{GBHA10}.  {Unlike the Bell scenario, the PM scenario
does not involve entanglement, and this makes it easier  to
 implement experimentally}~\cite{HGM+12,ABCB12,LBL+15}. {In {the PM scenario}, one party, the {{\em
preparer}}, prepares one of finitely many quantum states, then the
other party, the {{\em measurer}}, performs one of finitely many
measurements on the state}, see {FIG. 1(b)}. The corresponding {set
of measurement statistics} is called a \emph{PM correlation}.
{Similar} to the Bell scenario, a very important and natural problem
is to lower bound the dimension of the {quantum system} required to
generate a given PM correlation. {For this, the approach of linear
dimension witness was generalized to the PM scenario
in~\cite{GBHA10},} \comment{it was shown that the idea of dimension
witness still applies to this setting,} where the {preparer and the
measurer} share classical public randomness. The case {where the
devices are independent} was considered in \cite{BQB14}, where {one
needs to use nonlinear dimension witnesses}.

\begin{figure}[tbh]
\centerline{\includegraphics[width=3.4in]{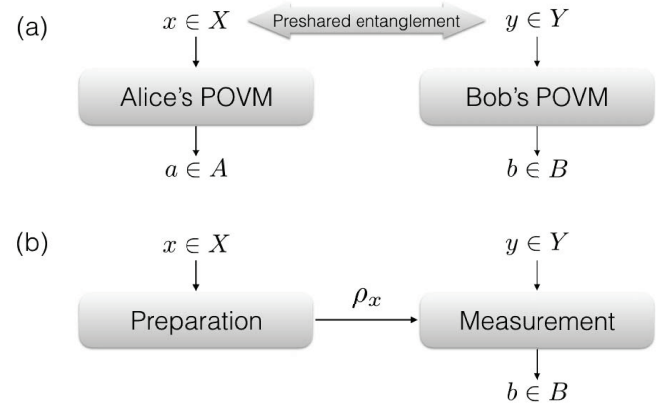}} \caption{(a) A
Bell scenario. (b) A prepare-and-measure scenario.}
\label{Bellsetting}
\end{figure}

Despite these encouraging {results}, dimension witnesses in the PM
scenario suffer from similar {drawbacks} as {those} for the Bell
scenario, which restricts their applicability. Indeed, some are very
specific, e.g., the dimension witnesses discussed in
\cite{GBHA10,BQB14} apply only for the case of binary measurements.
{Consequently, it is highly desirable to identify a lower bound for
the PM scenario, analogous to {that in} \cite{VSW15}, that is
applicable to PM correlations with {arbitrary} 
 parameters.
In this work, we provide such a new lower bound. To achieve this, we
first transfer the target PM scenario to a corresponding Bell
scenario, and then apply the bound given in \cite{VSW15}, leading to
{a new lower bound for PM correlations}. We show that the new lower
bound {performs very well} on some interesting applications, e.g.
quantum random access coding, and it also gives new insights for the
concept of dimension witness.
{Specifically}, 
 we
 show that the dimension witness provided in \cite{BNV13} 
{ can be obtained} as a direct consequence of {our new lower bound}.
{Furthermore}, we also use {our lower bound}  to prove that the sets
of {restricted-dimensional}  PM
correlations 
are not always convex.


{\em Scenarios.} A two-party Bell {scenario consists} of two
parties, Alice and Bob, that are in separate locations, and share a
quantum state $\rho$ acting on  $\C^{d_1} \otimes \C^{d_2}$. Alice
and Bob each have a (local) measurement apparatus acting on their
respective subsystems, see FIG. 1(a). A Bell correlation $r$ is the
collection of the joint conditional probabilities $r(a,b|x,y)$,
{i.e.,} the probability Alice and Bob {get}
 output $(a,b) \in A \times
B$ when they {use} measurement settings $(x,y) \in X \times Y$. In
\cite{VSW15}, it was {shown} that for a given Bell correlation
$r=r(a,b|x,y)$, both $d_1$ and $d_2$ are lower bounded by the
following two quantities:

\begin{small}
\begin{eqnarray}
& & \!\!\!\! \max_{y, y'}  \Big{(}
\displaystyle\sum_{b,b'}  \min_{x}  \Big{(}  \sum_{a}
\sqrt{r(a,b | x,y)} \sqrt{r(a,b' | x,y')} \Big{)}^2
\Big{)}^{-1}, \label{eq:lbound1} \\
& & \!\!\!\! \max_{x,x'}  \Big{(}
\displaystyle\sum_{a,a'} \min_{y} \Big{(}
 \sum_{b} \sqrt{r(a,b| x,y)} \sqrt{r(a',b| x',y)}
\Big{)}^2 \Big{)}^{-1}. \label{eq:lbound2}
\end{eqnarray}
\end{small}

As mentioned above, a PM {scenario} has one party preparing a
quantum state and the other measuring it, thus the outcome
probabilities are only seen on one side. {The preparer can generate
one out of  $N$ possible states, denoted by $\rho_x$,} where $x \in
X = \{1,2,...,N\}$. The measurer can choose one of $M$ different
measurements to perform, indexed by $y \in Y = \{ 1, \ldots, M \}$.
Each measurement $y$ {consists} of the operators $\{\Pi_b^y\}$,
where $b \in B = \{ 1, \ldots, K \}$ denotes the measurement
outcome. The probability of getting outcome $b$ when measurement $y$
is performed on quantum state $\rho_x$ can thus be expressed as
$p(b|x,y)=\Tr(\rho_x \Pi_b^y)$.

In this paper, we focus on the following problem: For a given {PM
correlation} $p$, what is the {smallest}
 dimension of a
quantum system
 that {is necessary to generate}
{it}? Throughout, we denote this quantity by $\mathcal{D} (p)$. Note
that one can always generate $p(b|x,y)$ if one chooses $\rho_x$ to
be the computational basis state ${\kb{x}}$
and $\Pi_{b}^y$ to be the diagonal {measurement} operator
$\sum_{z\in X} p(b|z,y) \kb{z}$. This proves that $\mathcal{D}
(p)\leq N$  for all PM correlations $p$. However, lower bounding
{$\mathcal{D}(p)$} {for a PM correlation} $p$  is a much more
interesting and challenging task.

{\em Deriving our lower bound.} {In this section, we will prove the
following theorem as the main result of the current paper.}

\medskip
\textbf{Theorem.} For any PM correlation $p$, we have that
$\mathcal{D}(p)$ is lower bounded by
\begin{equation} \label{thm:main}
\bigg{(} \sum_{x, x'} q_{x} q_{x'} \min_y \Big{(} \sum_{b}
\sqrt{{p(b|x,y)}} \sqrt{ {p(b|x',y)}} \Big{)}^2 \bigg{)}^{-1}
\end{equation}
for \emph{any} probability distribution {$q=(q_x)$} over $X$.

{\textbf{Proof}: Let ${d = \mathcal{D}(p)}$. Suppose $p$ can be
realized by the quantum states {$\{\rho_x\}_x $ and the measurements
$\{ \Pi_b^y \}_b$ acting on $\C^d$}. Consider the following Bell
scenario where Alice and Bob share the state
\begin{equation*}
\rho \equiv \sum_{x} q_x \kb{x} \otimes \rho_x
\end{equation*}
acting on $ \C^N \otimes \C^d$.} Here, $\{ \ket{x} \}_x$ is the
computational basis of $\C^N$, and $q$ is some fixed probability
distribution over~$X$. Furthermore, the two parties perform the
following measurements: Alice measures the subsystem on $\C^N$ in
the computational basis and Bob measures in the same way as he does
in the PM scenario (so the measurement outcome sets of Alice and Bob
are $X$ and $B$ respectively). See FIG. 2.

\begin{figure}[htbp]
   \centering
   \includegraphics[width=3.4in]{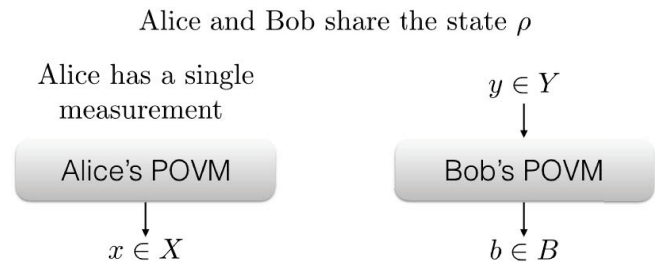}
   \caption{{Transforming} a PM scenario to a Bell scenario.}
   \label{fig:Bell}
\end{figure}

The Bell correlation corresponding to the strategy described above
is given by $r(x,b|y) = q_x \, {p(b|x,y)}$. Applying the lower
bounds \eqref{eq:lbound1} and \eqref{eq:lbound2} to the Bell
correlation~$r$, we get two lower bounds on $\mathcal{D}(p)$. {It
can be shown that one of the bounds always dominates the other, so
in the theorem we only give the larger of the two, which completes
the proof.}

{Some remarks on the lower bound \eqref{thm:main} are in order}.
Obviously, since $\mathcal{D}(p)$ is integral, we can round the quantity \eqref{thm:main} up if it is not an integer. Also one would naturally want to {identify} 
a probability distribution $q$ {that}
maximizes \eqref{thm:main}.
{In general, finding the  probability distribution {maximizing} \eqref{thm:main}} 
{corresponds to minimizing} a non-convex quadratic function over the
set of probability vectors (called the simplex). It is known that
solving such a problem is NP-hard~\cite{MS}. However, any
probability vector $q$ does yield a lower bound on
$\mathcal{D}(p)$. In this work, we mostly consider the uniform distribution, i.e., $q_x= 1 / N$ for all $x\in X$. 
{As it turns out this simple choice is} often sufficient to give tight bounds. Later we give an 
application where one can analytically prove this  is optimal.

{Lastly,}
whenever we have at most four possible preparations, i.e., $N\le 4$,
{there exists a tractable algorithm to find the distribution $q$
that maximizes \eqref{thm:main}.} {Specifically, in this case the
problem amounts to solving a semidefinite program, and this can be
done in polynomial time~\cite{Boyd}.}
For this, we use the fact that quadratic optimization over the
simplex can be expressed as a linear conic programming problem over
the cone of completely  positive matrices \cite{BDD+00}. An
$n$-by-$n$ matrix is {\em completely positive} if it can be written
as $\sum_{i=1}^k x_i x_i^{\top}$ (for some $k\ge 1$) where $x_i\in
\mathbb{R}^n_+$.
However, for $n\le 4$, an $n$-by-$n$ matrix is completely positive if and only if it is positive semidefinite and has nonnegative entries. 

{Before discussing applications of our lower bound, we first show  it can be 
tight, even when $q$ is  uniform.} Consider
the toy example where $X = \{ 0, 1 \}^M$, $Y = \{ 1, \ldots, M \}$,
$B = \{ 0, 1 \}$, and the PM correlation given~by
\begin{equation}
p(b|x,y) \equiv \delta_{b, {x_y}} \label{toy0}
\end{equation}
where $\delta$ is the Kronecker delta function and $x_y$ denotes the $y$'th bit of the bitstring
$x\in \{ 0, 1 \}^M$.
Intuitively, this means that given an encoding of the bitstring $x$,
 measurement can pick out {any} of the bits perfectly, thus can discriminate the quantum states received perfectly.
{In this case, our lower bound \eqref{thm:main} yields
$\mathcal{D}(p) \geq 2^M = N$, when $q$ is uniform. This is tight
since $\mathcal{D}(p) \leq N$ for any PM correlation.}

In the toy example above, we see that there is no way to \emph{compress} the states, that is, to reduce the dimension below the trivial bound of $N$. It turns out that there is a sufficient condition which follows easily from our main theorem. Since it is always true that
\begin{equation}
\sum_{x} \min_y \Big{(} \sum_{b}
\sqrt{{p(b|x,y)}} \sqrt{ {p(b|x,y)}} \Big{)}^2  = N,
\end{equation}
by setting $q$ to be uniform and applying our lower bound
\eqref{thm:main}, we get the following sufficient condition for {the
impossibility of quantum compressibility,}
 i.e., ${\mathcal{D}(p) = N}$:
\begin{equation}
\label{compress}
\forall x \neq x', \exists y, \forall b \text{ it holds that } p(b|x,y) \, p(b|x',y) = 0.
\end{equation}
If \eqref{compress} holds, $p$ cannot be (quantumly) compressed.

{\em New insights for dimension witness.} We now {introduce} two
examples to show that our lower bound provides new insights into the
concept of dimension witness.

We have mentioned that when the {preparer} and the measurer in a PM
scenario are independent, nonlinear dimension witnesses have been
proposed~\cite{BQB14,LBL+15}. {Suppose we have a PM correlation $p$
in the setting of $X = \{0,1,2,3\}$, {$Y=\{0,1\}$}, and $B=
\{0,1\}$. Then the determinant of the following $2 \times 2$ matrix
\[ W_2 = \begin{bmatrix}
p(0|0,0)-p(0|1,0) & p(0|2,0)-p(0|3,0)\\
p(0|0,1)-p(0|1,1) & p(0|2,1)-p(0|3,1)
\end{bmatrix}, \]
is a nonlinear dimension witness~\cite{BQB14}. In \cite{BQB14} it
was pointed out that if {${\rm det}(W_2)=2$} 
(the largest value possible), then $\mathcal{D}(p) \geq 4$. It can
easily be shown that this also follows from our sufficient condition
for non-compressibility. This is because when the determinant is
$2$, all the entries of $W_2$ must be $1$ or $-1$.} Then one can
check that the conditions \eqref{compress} are met, and we get
$\mathcal{D}(p) = 4$.

{For our second example, we recover the dimension witness in \cite{BNV13}.}
Specifically, we show something slightly more general, that for {$K=2$} and any $M$, $N$, we have that
\begin{equation} \label{binarybound}
\sum_{x,x'}\max_y\left|p(1|x,y)-p(1|x',y)\right|^2\leq \Big{(}1-\frac{1}{d} \Big{)} N^2
\end{equation}
is a quadratic  dimension witness for 
$d<N$. This can be easily shown by applying the Fuchs-van de Graaf
inequalities~\cite{FG99} to our lower  bound \eqref{thm:main} (with
$q$ uniform).

Note that in \cite{BNV13}, the dimension witness \eqref{binarybound}
was  derived in the special case   $M=N(N-1)/2$. In this case, the
{measurements $y$ in \eqref{binarybound} are {fixed and} labelled by
$y=(x,x')$, $x > x'$}. It turns out that in this case,
\eqref{binarybound} can be tight~\cite{BNV13}.

{\em {Relation to the Positive Semidefinite rank (PSD-rank)}.} We
first consider the case $M=1$, i.e., the measurer has only one
choice of measurement. {For this, we introduce the $N$-by-$K$  {row
stochastic} matrix $Q$
with entries given by $p(b|x)$. In this case, {it is known}
\cite{LWdW14}  that $\mathcal{D}(p)$ is {equal to the} {\em
PSD-rank} of $Q$, denoted $\rank_{psd}(Q)$, defined as the least
integer $c \ge 1$ such that there exist ${c \times c}$ {positive
semidefinite matrices} matrices $E_1, \ldots, E_N, F_1, \ldots, F_K$
satisfying $Q_{x,b} = \Tr(E_x F_b), \textup{ for all } x, b$. The
PSD-rank is an important quantity in computer science and
mathematical~optimization~\cite{FMP+12,GPT13,LSR14}.

We can expand this idea to more measurements in a few ways. First,
in \cite{LWdW14} it was shown that $\rank_{psd}(Q) \geq \sum_b
\max_x \, Q_{x,b}$ (this holds for any entry-wise nonnegative matrix
Q).  {Applying this to each $y$ individually, we  get}
\begin{equation}
\mathcal{D}(p) \ge \max_y \sum_b \max_x \, p(b|x,y), \label{PSDR}
\end{equation}
which was also {observed} in \cite{SH14}.
Notice that the {lower bound \eqref{PSDR}} 
is always upper bounded by $K$. Thus, our lower bound
\eqref{thm:main} can vastly outperform this bound as can be seen by
the PM correlation \eqref{toy0}. {Second, we can consider the matrix
$Q$ to the $KM$-by-$N$ matrix $Q'$
with entries given by {$Q'_{(y,b),x}=p(b|x,y)$}. By definition, it
follows that $\mathcal{D}(p) \geq \rank_{psd}(Q')$. Then {using the
technique to lower bound PSD-rank introduced in \cite{LWdW14}, we
can recover \eqref{thm:main}, giving an alternative proof of our
main result.}

{\em Quantum random access codes.} We now consider the PM
correlations arising from an  information task known as \emph{random
access coding}. The goal here is to encode $M$ bits into a quantum
state of hopefully small dimension such that the measurer can choose
\emph{any of the $M$ bits} to learn with high probability. Moreover,
consider a PM {scenario} with $X = \{ 0, 1 \}^M$, $Y = \{ 1, \ldots,
M \}$, and $B = \{ 0, 1 \}$ where the PM correlation is given by
\begin{equation}
p(b|x,y) \equiv \left\{
\begin{array}{cl}
\beta & \text{ if } b = x_y, \\
1 - \beta & \text{ if } b \neq x_y.
\end{array} \right.
\label{toy}
\end{equation}
Here $\beta \geq 1/2$ is the success probability of learning the
$y$'th bit correctly. There are well-known examples where a single
qubit can encode $2$ bits with $\beta \approx 0.8536$ (see
\cite{BBBW83, ANTV99}) or $3$ bits with $\beta \approx 0.7887$ (see
\cite{Chuang, ANTV02}). {We see that on both these examples, the
bound \eqref{thm:main} is tight (when $q$ is uniform). In fact, it
can be shown that $q$ being uniform is the optimal choice when
applying our lower bound \eqref{thm:main} for this case of random
access codes~\cite{footnote1}.}

We now ask the question of how the dimension is affected when
$\beta$ changes. We compare our bound to Nayak's bound~\cite{Nay99},
which is essentially optimal and states that a random access code
requires at least $(1 - H(\beta)) M$ qubits, where $H$ is the binary
entropy function $H(\beta) \equiv - \beta  \log_2(\beta) - (1 -
\beta) \log_2(1 - \beta)$. In other words, Nayak's bound can be
expressed as ${\mathcal{D}(p) \geq \lceil 2^{(1 - H(\beta)) M}
\rceil}$.

{For small values of $M$, our bound behaves very well by being quite
close to Nayak's bound. For example, for $M = 2$, Nayak's bound
beats our bound for $\beta \in (0.8900, 0.9083) \cup (0.9674,
0.9714)$. For all the other values of $\beta$ we get the same bound.
Thus, our bound performs very well and is close to optimal in this
setting.}

However, Nayak's bound is concerned with the \emph{worst case}
probability of correctly decoding a bit. {Therefore}, one can easily
construct other PM correlations where our lower bound is greater.
For example, {if we alter only a few of the states in a random
access code such that for some $x$, some of the bits are decoded
with a very small success probability, Nayak's bound will approach
$1$ (as the binary entropy will approach $1$). On the other hand,
our bound can still be large as it deals with all of the outcome
probabilities independently.}

{\em {Witnessing the non-convexity of {restricted-dimensional} PM
correlations.}} {We now study the sets of
\emph{{restricted-dimensional}} PM correlations, that is, {the sets
${D_{d} \equiv \{ p : \mathcal{D}(p) \leq d \}}$}, for some fixed
integer $d \ge 1$. It was first pointed out in \cite{GBHA10} that
there exist choices for $d$ for which ${D}_d$ is not convex. We now
show that this  can {be proved easily} using the lower bound
\eqref{thm:main}.}

{For this}, consider the PM scenario with $X = \{ 0, 1 \}^2$, $Y =
\{ 1, 2 \}$, and $B = \{ 0, 1, 2 \}$. {For $i=1,2$
 define}
\begin{equation}
p_i (b|(x_1, x_2), y) \equiv \left\{
\begin{array}{cl}
\delta_{b, x_i} & \text{ if } y = i \\
\delta_{b, 2} & \text{ if } y \neq i.
\end{array} \right.
\end{equation}
{Intuitively, this is similar to our toy example \eqref{toy0} where
one wants to perfectly decode one of two bits, except that the
measurer can output ``I am not certain'' (as indicated by the
outcome ``$2$''). Then $p_i$ is the PM correlation where the
preparer chooses two bits $(x_0, x_1)$, sends $x_i$, and the
measurer learns and outputs {$x_i$} if and only if $y = i$ (and
outputs ``$2$'' otherwise). For this reason, it is clear that each
$p_i$ is in $D_2$.}

{The convex combination is more than just sending a random choice of
$x_1$ or $x_2$ as the measurer's POVMs need to be taken into
account. It turns out that any qubit encoding is not possible as our
lower bound \eqref{thm:main} applied to $\frac{1}{2} p_1 +
\frac{1}{2} p_2$ (with $q$ uniform) is equal to $16/7 > 2$.
Therefore, $\frac{1}{2} p_1 + \frac{1}{2} p_2 \not\in D_2$,
witnessing that $D_2$ is not convex.}

{\em Conclusions.} In this work we derived a lower bound for the
dimension of any quantum system that can produce a given PM
correlation, which is applicable {to} any {choice of} parameters,
and has nontrivial applications. {Specifically}, we showed that our
lower bound provides new insights for the notion of dimension
witness. We also used the bound to prove that the set of
{restricted-dimension} PM correlations is not always convex. Due to
the generality of the PM scenario we believe that our lower bound
will lead to more nontrivial applications, and {will} provide new
{insights} {into} the study of device-independent quantum processing
tasks.

\begin{acknowledgments}
J.S. is supported in part by NSERC Canada. A.V. and Z.W. are
supported by the Singapore National Research Foundation under NRF RF
Award No.~NRF-NRFF2013-13.
{Research at the Centre for Quantum Technologies at the National University of Singapore is partially funded by the Singapore Ministry of Education and the National Research Foundation, also through the Tier 3 Grant ``Random
numbers from quantum processes,'' (MOE2012-T3-1-009).}
\end{acknowledgments}


\begin{thebibliography}{9}
\bibitem{footnote1}
{The quadratic optimization problem of finding the distribution $q$
maximizing \eqref{thm:main} turns out to be {convex in this case}
and furthermore, using the Karush-Kuhn-Tucker conditions (e.g. see
\cite{Boyd}), it can be verified that the optimal $q$ is uniform.}
\end{thebibliography}
\end{document}